\documentclass[conference]{IEEEtran}
\IEEEoverridecommandlockouts
\usepackage{cite}
\usepackage{amsmath,amssymb,amsfonts}

\usepackage{graphicx}
\usepackage{textcomp}
\usepackage{xcolor}
\usepackage{array, caption, tabularx,  ragged2e,  booktabs}
\usepackage{enumitem}
\usepackage{arabtex}
\usepackage{tabularray}
\usepackage{comment}
\usepackage{algorithm}
\usepackage{algpseudocode}

\usepackage[]{url}

\def\BibTeX{{\rm B\kern-.05em{\sc i\kern-.025em b}\kern-.08em
    T\kern-.1667em\lower.7ex\hbox{E}\kern-.125emX}}
\begin{document}

\title{Design and Implementation of a Controlled Ransomware Framework for Educational Purposes Using Flutter Cryptographic APIs on Desktop PCs and Android Devices\\
} 

\author{
James Gu, Ahmed Sartaj, Mohammed Akram Taher Khan and Rashid Hussain Khokhar \\
School of Computer Science \& Technology, Algoma University, Sault Ste. Marie, ON, Canada \\ Emails: 
\{jgu, asartaj, mtaherkhan, rashid.khokhar\}@algomau.ca
}

\maketitle

\begin{abstract}

This study focuses on the creation and implementation of ransomware for educational purposes that leverages Python's native cryptographic APIs in a controlled environment. Additionally, an Android version of the framework is implemented using Flutter and Dart. For both versions, open-source cryptographic libraries are utilized. With this framework, researchers can systematically explore the functionalities of ransomware, including file encryption processes, cryptographic key management, and victim interaction dynamics. To ensure safe experimentation, multiple safeguards are incorporated, such as the ability to restrict the encryption process to a specific directory, providing the RSA private key for immediate decryption, and narrowing the scope of targetable files to a carefully curated list (.txt, .jpg, .csv, .doc). This paper draws inspiration from the infamous WannaCry ransomware and aims to simulate its behaviour on Android devices. By making the codebase open-source, it enables users to study, modify, and extend the program for pedagogical purposes and offers a hands-on tool that can be used to train the next generation of cybersecurity professionals.

\end{abstract}

\begin{IEEEkeywords}
Ransomware Simulation, RSA, AES, Flutter, Python, Dart, Android Port, Desktop
\end{IEEEkeywords}

\section{Introduction}
Ransomware remains a formidable threat in the cybersecurity landscape, with global damages projected to exceed 20 billion USD annually, according to Beaman et al.~\cite{BEAMAN2021102490}. This underscores the growing sophistication and prevalence of ransomware attacks, which exploit system vulnerabilities to encrypt data and demand payment for its release. A prominent example is the WannaCry attack, which occurred on May 12, 2017~\cite{8323682}, and rapidly spread to infect over 200,000 computer systems across 150 countries. It affected many critical systems, most notably the United Kingdom’s National Health Service (NHS), where it locked access to patient records, disrupted medical procedures, and delayed emergency care, resulting in estimated damages exceeding £92 million~\cite{natureRetrospectiveImpact}.

Analyzing real ransomware requires isolated environments, such as virtual machines or sandboxes, to prevent accidental damage to systems or networks. However, this process is resource-intensive and often inaccessible to many learners. Our framework addresses this gap by constructing a benign ransomware simulation that replicates core functionalities such as file encryption, key exchange, and ransom demands, in a controlled and reversible manner. By providing a safe alternative, we aim to enhance cybersecurity education by equipping learners with practical insights into ransomware mechanics and mitigation strategies, while also enabling the development of robust defenses against real-world attacks.\\

The contributions of the paper are:
\begin{itemize}
    \item \textit{An Open-Source Framework for Ransomware:} An architecture for ransomware that's functional on both desktop and Android environments, with an implementation which is openly accessible with comprehensive documentation\footnote{The GitHub repository containing the source code is available at: \url{https://github.com/jgualgoma/ransomware_demo/}} . \\

    \item \textit{Survey of Malware and Ransomware Tools:} The paper provides a review of widely available open-source packages and libraries for cryptography. We aim to showcase areas of potential exploitation by malicious actors and areas where developers can implement stronger security measures in their applications.\\ 

    \item \textit{Recursive File Encryption and Decryption Algorithm:} A novel recursive encryption and decryption algorithm is introduced. The algorithm employs a hybrid encryption approach, using AES for efficient file encryption and RSA to secure the AES keys, mirroring real-world examples, such as WannaCry.
    
\end{itemize}
    
The rest of the paper is organized as follows: In the \textit{Literature Review} section, we discuss the evolution of ransomware, cryptographic techniques, anonymous payment mechanisms, notable ransomware attacks, and ransomware statistics. The \textit{System Overview and Scope} section outlines the framework and overall architecture, such as the encryption mechanisms employed. In the \textit{System Design and Implementation} section, we describe the technology stacks and libraries used to implement the ransomware, along with algorithms employed. The \textit{Challenges and Design Decisions} section discusses and justifies key design decisions. In the \textit{Evaluation} section, we present the testing and validation conducted for both Android and Desktop implementations. Finally, the \textit{Conclusion} section summarizes our findings.

\section{Literature Review}

The first ransomware dates back to the AIDS Trojan in 1989, created by Dr. Joseph Popp. It used symmetric encryption to scramble file names on floppy disks and demanded \$189 via mail~\cite{young1996cryptovirology}. Though primitive, it introduced the concept of data extortion through encryption, laying the groundwork for future developments. In the early 2000s, GPCode~\cite{nakamoto2008bitcoin} marked a shift by employing more advanced RSA-1024 encryption, albeit with several flaws that allowed data recovery. Scareware variants like Reveton~\cite{liska2016ransomware} later emerged, locking screens with fraudulent law enforcement notices and demanding fines, leveraging psychological pressure and deception. 

Bitcoin further revolutionized ransomware by providing an anonymous payment mechanism, fueling its proliferation. CryptoLocker~\cite{fiore2023security}, utilizing an RSA-2048 and AES-256 hybrid model, spread through phishing emails via malicious attachments, extorting \$300-\$700 per victim and amassing a massive \$27 million in bitcoin ransom before its takedown in 2014 via Operation Tovar~\cite{liska2016ransomware}. 

Fiore et al.~\cite{fiore2023security} described how the Locky ransomware targeted healthcare institutions with phishing campaigns, encrypting patient records and demanding ransoms in Bitcoin, exploiting the sector’s need for immediate data access. Ryuk~\cite{fayi2018petya}, linked to North Korea's Lazarus Group and similar to WannaCry, focused on high-value enterprise targets, demanding up to \$5 million per ransom. Petya and its derivative NotPetya~\cite{fayi2018petya} leaned toward data destruction, encrypting master boot records to render systems unbootable, costing companies like Maersk \$300 million in damages and blurring the line between ransomware and cyberwarfare. It was deployed as a geopolitical weapon, particularly in countries like Ukraine.

The WannaCry outbreak of May 12, 2017, infected over 200,000 systems across 150 countries by exploiting a Windows SMB vulnerability (CVE-2017-0144) through the EternalBlue exploit~\cite{8323682}. Its hybrid model, which used AES-128 for file encryption and RSA-2048 for key encryption, combined with worm-like propagation, caused widespread disruption, notably costing the NHS £92 million~\cite{natureRetrospectiveImpact}.

The WannaCry architecture remains relevant as researchers continue to develop new ways to protect systems from ransomware. In one study, researchers emulated six prominent ransomware architectures, including WannaCry, with high accuracy to analyze their behaviour~\cite{10.1145/3688351.3689163}. This underscores the importance of creating ransomware simulations and studying their attack vectors, as many modern variants are descendants of early ransomware like WannaCry.

Furthermore, ransomware has increased its capabilities in recent times. In one instance, researchers developed a File System Access API (FSA) and WebAssembly (Wasm) based infection vector, enabling ransomware to propagate through browsers~\cite{10.1145/3708514}. This highlights the need to study different attack vectors of modern ransomware, as we aim to do on the Android mobile platform. Other studies have also analyzed ransomware and its detection using machine learning and deep learning techniques~\cite{KRITIKA2025100078}. This makes simulations like ours especially important for safely and effectively training and developing AI-powered detection systems.

\section{System Overview and Scope}

\subsection{Desktop PC Python Implementation}

This paper focuses on the development of a ransomware simulation within a controlled framework. It encrypts files located in a specified \textit{files} directory using a hybrid cryptographic approach that combines RSA (Rivest-Shamir-Adleman) and AES (Advanced Encryption Standard), similar to ransomware such as WannaCry. The simulation features a graphical user interface, implemented via Python’s \texttt{tkinter}\footnote{Used to design the ransomware’s user interface. Available at: \url{https://docs.python.org/3/library/tkinter.html}} package, to deliver the ransom demand and provide an immersive experience. The code is compiled into a standalone executable compatible with Windows, macOS, or Linux systems, ensuring broad accessibility for educational use. 

Key aspects covered within this scope include the implementation of recursive file encryption to simulate ransomware’s systematic targeting, hybrid key management to demonstrate ransom exchange, and a UI for educational purposes. For safety, the program restricts its operations to a predefined set of file extensions (.txt, .jpg, .csv, .doc), and provides the RSA private key as a \texttt{key.pem} file for immediate decryption, ensuring all actions are fully reversible. This implementation explicitly excludes advanced malicious features, including network propagation mechanisms, such as worm-like spreading via SMB ports, as seen in WannaCry; exploitation of operating system vulnerabilities, such as EternalBlue; and persistence techniques that would enable the program to persist through system reboots. It does not aim to evade antivirus detection, bypass security software, or simulate real-world attack vectors beyond the fundamental processes of encryption and notification. Instead, it concentrates on delivering an educational experience that illustrates ransomware operations within a controlled, ethical framework.

\subsection{Mobile Android Implementation}
Additionally, an Android app is developed using the Flutter/Dart framework. This Android port is a one-to-one translation of the Python desktop application, but must employ a new encryption package and a distinct file system. This port is packaged as an APK that runs as a standalone Android app, and is tested on API 36, the latest Android version available. 

\subsubsection{File Access Systems}
One major restriction of the mobile Android port is its limited ability to access user files. Direct file access has been restricted, and scoped storage has been enforced since API 29. The Google Play store also rejects any apps that still request direct file access permissions~\cite{Android1}. As a result, each application running on Android OS has its own isolated file storage. When accessing files from common directories, such as camera pictures, the OS instead caches the file, creating a temporary copy in a temporary directory for the application to access. This poses a particular problem for ransomware, as direct read and write access to user files is crucial.

\subsubsection{Storage Access Framework}
The Storage Access Framework is a complementary framework for file access based on user permissions~\cite{Android2}. By using this framework and its associated APIs, an application can retain permissions to read from and write to storage directly if a user grants access to a specific file or folder. This means that, through social engineering or user carelessness, a ransomware app that receives explicit permissions could function on a mobile device as effectively as it would on a desktop.

\subsubsection{Dart Packages}
The main encryption package used in the Android port is the \texttt{encrypt} package. It contains the Fernet submodule, which is equivalent to the Python cryptography Fernet submodule and is used in the AES encryption scheme. It also contains an RSA submodule, which is used to encrypt the randomly generated AES keys. The main differentiator between the Android and desktop versions is the inclusion of the \texttt{saf\_stream} \footnote{Dart Package for reading and writing using the Storage Access Framework APIs for Android OS. Available at: \url{https://pub.dev/packages/saf_stream}} and \texttt{saf\_util} \footnote{Dart Package for navigating the file system using the Storage Access Framework APIs. Available at: \url{https://pub.dev/packages/saf_util}} packages. These Dart-specific packages utilize the SAF APIs to allow direct file access when given the proper permissions, enabling the ransomware to encrypt files inside the chosen directory. 

\section{System Design and Implementation}
This paper presents the design and implementation of a ransomware simulation tailored for educational purposes, contributing a unique and valuable resource to the field of cybersecurity. The encryption and decryption logic is developed using the Python \texttt{cryptography} package to integrate RSA and AES algorithms. These methods mirror techniques employed by real-world ransomware variants, such as WannaCry, which serves as a starting point for architectural insights and design principles. A graphical ransom interface is created using the \texttt{tkinter} library to simulate victim interaction, visually demonstrating the ransom demand process and decryption mechanism. The simulation is made portable across multiple operating systems by utilizing the \texttt{pyinstaller} tool to generate standalone executables, ensuring its versatility in diverse educational settings.

The value of our work lies in its ability to raise awareness about ransomware threats while providing hands-on learning in a risk-free environment. Unlike WannaCry, a malicious, Windows-specific ransomware that exploited the SMBv1 vulnerability for propagation and sought financial gain, our simulation is benign, platform-agnostic, and designed for educational purposes. This key distinction enables students and researchers to dissect ransomware mechanics, such as the hybrid encryption model and user interaction strategies, understand cryptographic principles, and explore mitigation techniques without the ethical dilemmas or technical hazards associated with live malware analysis. Our contribution stands out by offering a practical, accessible tool that contrasts sharply with WannaCry’s destructive purpose, enhancing cybersecurity education and preparedness across academic and professional communities.

The main contribution of the Android port is identifying the ways in which packages and APIs of the Android operating system can be exploited through social engineering or user negligence to grant ransomware access to read and write permissions. The port also demonstrates how commonly used packages, such as the Dart/Flutter framework, can be exploited to produce malware.

\subsection{Technology Stack}
The Flutter and Dart technology stack is chosen for its performance and flexibility in mobile application development, as well as its cross-platform capabilities for both web and mobile. One of the goals is to evaluate the difficulty of creating ransomware from scratch using limited tooling. The only additional packages required beyond the base Dart installation are the cryptography package and those necessary for file access. The Dart/Flutter script can then be compiled into a native executable for the Android platform, allowing it to run independently without requiring Dart SDK support. This makes the ransomware highly efficient, capable of compiling into a working program on any Android device with minimal compatibility issues.

\subsection{Cryptographic Packages}
The Dart cryptography package is used for all key generation and file encryption. Initially, a public/private key pair is generated using the RSA submodule of the package. The generated code is pushed to a source code repository and is commented out. Since only one key pair is needed, the private key is saved as a \texttt{PEM} file and used for decrypting files, while the public key is stored as a byte string within the source code. This does not pose a security risk, as the public key can only encrypt files. Therefore, even if the byte string is discovered during reverse engineering, it cannot be used to decrypt the affected files.

\subsection{Cryptographic Algorithms}

\begin{algorithm}
  \caption{Recursive Encryption Algorithm}
  \begin{algorithmic}[1]

    \Require PU\_$Key_{(RSA)}$
    \Require extension\_list
    
    \Procedure{recursive\_encrypt}{$DIR$}

      \For{\texttt{file in $DIR$}}
     
        \If{file is directory}
            \State \texttt{recursive\_encrypt(file)}
        \ElsIf{file in extension\_list}
            \State \texttt{$Key_{(AES)}$= GenAES()}
            \State \texttt{$E_f$= Fernet( $Key_{(AES)}$, file)}
            \State \texttt{$K_E$= RSA( PU\_$Key_{(RSA)}$, $Key_{(AES)}$)}
            \State \texttt{dict[$K_E$]= $E_f$.name }
            \State \texttt{dict.writeToDisk()}
        
        \EndIf
      \EndFor
    \EndProcedure
  \end{algorithmic}
\end{algorithm}

\begin{algorithm}
  \caption{Decryption Algorithm}
  \begin{algorithmic}[1]
    
    \Require keys\_datafile
    
    \Procedure{decrypt}{$keys\_datafile$}

         \If{PR\_$Key_{(RSA)}$ exists}

            \For{\texttt{$K_E$, $E_f$ in $keys\_datafile$}}
     
                \State \texttt{$Key_{(AES)}$=RSA(PR\_$Key_{(RSA)}$, $K_E$)}

                 \State \texttt{file= Fernet($Key_{(AES)}$, $E_f$)}
                                  
            \EndFor
        \EndIf
      
    \EndProcedure
  \end{algorithmic}
\end{algorithm}

The two primary cryptographic algorithms employed are RSA (Rivest–Shamir–Adleman), an asymmetric public/private key block cipher, and AES (Advanced Encryption Standard), a symmetric key block cipher. The RSA algorithm relies on mathematics and number theory to generate a key pair in which the public key is used to encrypt data, and only the corresponding private key can decrypt it. The AES algorithm, by contrast, uses the same key for both encryption and decryption and operates as a substitution–permutation cipher. When using a 128-bit key, AES executes significantly faster than RSA on modern systems. 

The goal of using RSA and AES in unison is to leverage the strengths of both algorithms. While RSA provides robust security due to its complex mathematical computations, it is inefficient for encrypting a large number of files. The AES algorithm, on the other hand, is much faster due to its permutation-substitution-based approach. Modern malware, such as WannaCry, has adopted a hybrid model in which the initial encryption of files is performed using the AES algorithm, with each file encrypted using a unique AES key. The keys are then encrypted using the public key hard-coded into the malware, which poses no issue for the attacker because only the private key held by the attacker can be used to decrypt the files.

A recursive encryption algorithm used to crawl through the file system and encrypt all files that have a matching extension defined in the extension list. This algorithm encrypts all identified files and stores an association between each encrypted file and its corresponding encrypted key so that the file can later be decrypted using its associated key. It uses GenAES to generate a random AES key, which the Fernet algorithm then uses to encrypt the files. The AES key is encrypted with a hard-coded RSA key, and the encrypted AES key is associated with the file name. This association is stored on disk. 

A custom Decryption algorithm is used to decrypt the user files. This algorithm executes only if a private key in the form of a \texttt{.PEM} file exists in the current working directory. This private key is provided to the victim after they have deposited the ransom into the specified Bitcoin wallet, as outlined in the ransom demand note. If the private key is available, the algorithm searches for a data file on disk which houses the encrypted AES key and its associated file. The RSA algorithm is then used to decrypt the encrypted AES key, and the resulting decrypted AES key is used to decrypt the associated file, thereby restoring access for the user. 

It should be noted that the payment process occurs outside the program, as the user is prompted to deposit the ransom into a Bitcoin wallet monitored by the attackers. Once the ransom has been received, the private key required for decryption is sent to the user. As shown in Algorithm 2, the algorithm checks for the presence of the private key, which it uses for decryption. In future updates, the ransomware may include a module to automatically download the private key from a remote server once the ransom has been paid. However, this feature is not essential to the current operation of the ransomware.

\subsection{Safety Controls via STRIDE modeling}
A full STRIDE analysis is presented in Table 1, where we summarize potential threats and mitigation strategies. The major areas we have identified include Spoofing, Tampering, and Repudiation. Bad actors may pretend to be the developers and distribute actual malware instead of the simulation. To mitigate this, users should obtain the code only from the official GitHub repository. To prevent tampering with the source code, commit permissions are restricted to verified contributors, thereby reducing the risk of unauthorized insertion of malicious code. For Repudiation, we provide a checksum so that users can verify it against their build to ensure that the compiled code originates from the authors. Minor threats include Information Disclosure, Denial of Service, and Elevation of Privilege. These can be mitigated through basic safety training and best practices,  as outlined in the mitigation strategies section of Table 1.

\begin{table*}
    \centering
    \begin{tabular}{|c|p{4cm}|p{4.5cm}|p{4.5cm}|c|}
    \hline
         \textbf{STRIDE Category} &\textbf{Threat}  & \textbf{Potential Impact } & \textbf{Mitigations}\\
\hline
         S–Spoofing& Impersonation of developer/tester or user  &An adversary may impersonate a developer to distribute suspicious and maliciously modified versions of this ransomware.  & Obtain the source code only from the official GitHub repository.  \\
         \hline
         T –Tampering& Modifying code or cryptographic elements  &  An adversary may alter the encryption logic or tamper with the public key. & Implement code integrity checks, such as hashing, and restrict access to the code repository. \\
         \hline
         R – Repudiation&Denial of actions performed & It may allow users or attackers to claim that they did not execute or modify the code.   &  Keep a record of all users who have accessed or forked the GitHub repository, and provide checksum for the codebase. \\
         \hline
         I – Information Disclosure&Leakage of encrypted data  & Users may use the provided RSA public-private key pair for unintended purposes.  & Remind users that the provided public key is for demonstration purposes only and should not be used in any other context.\\
         \hline
         D – Denial of Service& Misusing encryption logic to corrupt files &  Users may inadvertently misuse the program and corrupt the file system. & The encryption logic is restricted to certain directories, so that even with permissions, the encryption simulation cannot operate outside its bounds.\\
         \hline
         E – Elevation of Privilege&  Unauthorized access to file system & The application requires elevated permissions on Android.
 & Revoke those permissions afterward, or use an emulator to prevent misuse of elevated permissions.\\
         \hline
    \end{tabular}
    \caption{STRIDE analysis and mitigation strategies}
    \label{tab:my_label}
\end{table*}

\begin{figure*}
    \centering
    \includegraphics[width=1\linewidth]{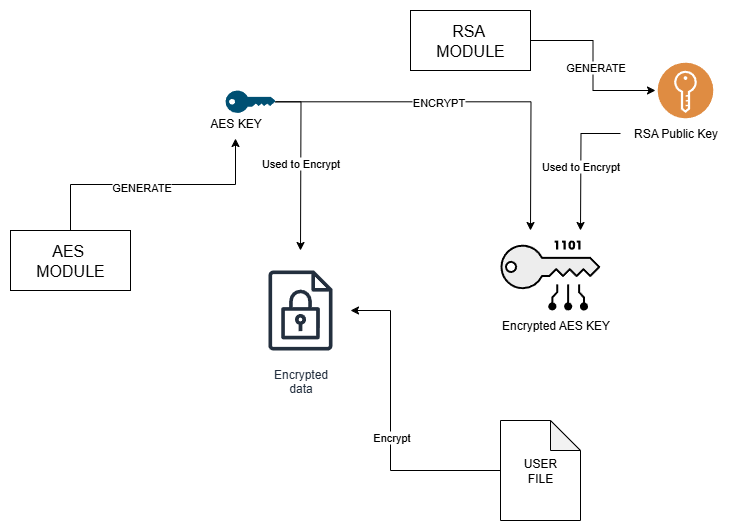}
    \caption{Workflow of the Ransomware Encryption}
    \label{fig:enter-label}
\end{figure*}

\section{Challenges and Design Decisions}

\subsection{Balancing Realism and Safety}
Replicating ransomware behavior poses a risk of unintended harm if the simulation’s scope is not adequately constrained, potentially encrypting critical files or spreading beyond intended boundaries. To address this, the simulation is designed to restrict encryption to a specific directory using “os.path” checks to validate file paths and ensuring operations remain within the user’s control. Furthermore, target extensions are limited to a whitelist (e.g., .txt, .jpg, .csv, .doc), preventing access to system files or sensitive data, unlike WannaCry’s broad targeting of user directories and critical infrastructure~\cite{8323682}. The objective is to simulate a realistic scenario where a system could be compromised by ransomware, while avoiding jeopardizing the safety and security of the host system.

\subsection{AES Key Management}
Following the blueprint of the WannaCry ransomware, each file is encrypted with its own AES key. As depicted in Figure 1, during encryption, each encrypted file must be associated with its corresponding AES key. This association is maintained using a \texttt{keys.dat} file, which serves as an intermediary structured as a dictionary that maps file names to their encrypted AES keys. To ensure key confidentiality, each AES key is encrypted using a public RSA key hard-coded into the ransomware's source code. This design ensures that the AES keys cannot be used to decrypt the corresponding files unless they are first decrypted using the correct RSA private key.

\section{Evaluation}

\subsection{Desktop Platform}
The desktop simulation is tested on Windows, macOS, and Linux using the \texttt{pyinstaller} package to create a standalone executable. The executable runs consistently across all platforms, ensuring cross-platform compatibility. 

The evaluation confirms the following:
\begin{itemize}
    \item Algorithm 1 successfully encrypts targeted files, while Algorithm 2 efficiently decrypts them, restoring files to their original state without data loss.
    \item The Tkinter-based ransom note interface effectively simulates victim interaction and demonstrates usability for educational purposes.
    \item Safety controls restrict encryption to the target directory and enforce a whitelist of file extensions, ensuring both reversibility and system safety.
    \item On a test system (Intel Core i7-12700H, 16 GB RAM, SSD), 100 files (500 MB; .txt, .jpg, .csv, .doc) were encrypted in 12.4 s and decrypted in 14.2 s.
    \item For Comparison to WannaCry, Kao et al.~\cite{8323682} indicate WannaCry encrypted at ~10 MB/s in 2017, taking ~50 seconds to encrypt 500 MB. This simulation’s ~40 MB/s rate is considereably faster, though modern ransomware may have improved during this time.
\end{itemize}

\subsection{Android Platform}
Testing of the Android port is conducted on a Google Pixel 9 emulated device using Android Studio, running Android 16. API 36, the latest API version for Android, is utilized in the tests. The ransomware APK successfully encrypts and decrypts user files (e.g., .txt, .jpg) when user permissions are granted, and it is able to decrypt the locked files once the private key is provided. The workflow of the Android port is the same as that of the desktop version, with the exception of an additional step where the user must first grant permissions to the app. It should be noted that performance testing could not be reliably conducted, as an Android emulator was used for testing and validation, making comparisons to real-life hardware unrealistic.

\section{Conclusion}
This paper successfully delivers a safe, functional ransomware simulation, and offers students and security researchers a practical means to explore cybersecurity threats without the ethical or technical risks associated with live malware~\cite{8323682}. 
This framework implementation enhances  understanding of ransomware threats while acting as a practical resource for academic and professionals.

Although security for Android OS is tightening, reliance on legacy systems, such as SAF, allow ransomware to exploit such vulnerabilities. While an Android app as described in this paper would not be accepted by the Google Play Store, illegal modification and distribution of mobile apps are common, and malware such as ransomware can easily be included with them, as outlined in this paper.

Overall, all versions of our ransomware implementations effectively leverage existing cryptographic APIs, showcasing the ease by which malware creation can occur using open source tooling.

\bibliographystyle{IEEEtran}


\end{document}